# Isospectral Potentials and Quarkonia Low Energy Spectra: A Possibility of simultaneous fit to Masses and leptonic decay widths".

Nitika Sharma, PK Chatley
*Department of Physics, National Institute of Technology*
*Jalandhar-144 011, India*

Avinash Sharma[*]
*School of Basic & Applied Sciences,*
*GGS Indraprastha University*
*Delhi-110 403, India*

Some of the non-relativistic QQ/bar interaction potentials and its isospectral partner-potentials have been employed to evaluate the masses and leptonic decay widths of Charmonium and Upsilon states. An attempt has been made to develop a prescription to provide a simultaneous fit to Masses and leptonic decay widths of Quarkonia. Preliminary calculation shows that the class of isospectral-potentials corresponding to: $V(r) = -(\alpha_s/r) + ar^n$ appears to yield the desired results. The various associated theoretical and phenomenological issues are discussed in light of recent data. The possible relation with the '*Wrinkled potentials*' is also discussed.

---

[*] [acsharma1956@gmail.com](mailto:acsharma1956@gmail.com)

In a recent paper Sharma et el.[1] have argued that an incompatibility exists while an attempt is made to fit simultaneously the two of the most well studied properties of the quarkonium, namely, masses and leptonic decay widths within an overall nonrelativistic potentials model framework.

In particular, power law potential of the form

$$V(R) = -\frac{\alpha_s}{R} + a R^n,$$

is employed to recalculate the masses and the leptonic decay widths of S-states of the charmonium and bottomonium by considering a and n as free parameters. It has been shown that the fitting of the masses prefers large n value ( n > 1) and smaller a value while the leptonic widths prefers smaller n values (n < 1) i.e. convex downward potentials. The ranges of n & a required for these physical quantities lie in the diagonally opposite corners of the a-n plane. The separation further increases in the presence of QCD corrections. It may be noted that the class of potentials represented by the n < 1 are concave downward and the one represented by n > 1 are convex downward. It has been observed that masses and leptonic decay widths tend to choose the exponents which represent different curvatures for the quark-antiquark potential. It has also been argued that the relativistic corrections also do not improve the situation.

The isospectral potentials[2] have identical energy levels except the ground state and may well have altogether different curvatures. The potentials with different curvatures shall have different *wave function overlaps at origin*.leading to different leptonic decay widths. This provides a framework to have a simultaneous fit to the mass spectrum and the leptonic widths data.

In the present work masses and the leptonic decays of the charmonium and bottomonium have been recalculated using the QQ/bar potential as V(r)= -(alphas/r)+ar^n and one of its typical isospectral counterpart potential. The preliminary results show that the hypothesis conjectured in above seems justified. The details of the formalism and the calculation with details results are presented elsewhere[3] .

## Acknowledgments

The authors wish to thank Profs. MM Gupta and RC Verma for useful discussions.

**References**

[1]    A.C. Sharma, R.C. Verma and M.P. Khanna, Ind. J. of Pure & appl. Phys. 36, (1998) 259.

[2]    T.D. Imbo and U. Sukhatme, Phys. Rev. Lett, 54 (1985) 2184; A. Khare & U Sukhtme, J. Phys. A22 ( 1989) 2847; WY et.al., J.Phys. 22(1989) L987 A. Khare & U. Sukhatme, Phys. Rev. A40, (1989) 6185, and references therein.

[3]    Neetika Sharma, P.K. Chatley and .A. C Sharma, Preprint no. NIT-J/Phys. 04 /2008; Submitted to Phys. Rev. D (2008)